\documentclass[12pt]{article}
\usepackage{latexsym}

\def\dag{\dagger} \def\pd{\partial} \def\pp{\prime} \def\a{\alpha} \def\b{\beta} \def\dl{\delta} \def\s{\sigma}  \def\eps{\epsilon} \def\veps{\varepsilon} \def\lam{\lambda} \def\Lam{\Lambda} \def\gm{\gamma}  \def\om{\omega} \def\Om{\Omega} \def\sq{\sqrt} \def\fr{\frac} \def\half{\frac{1}{2}}

\def\hg{{\hat g}}  \def\nb{\nabla} \def\hnb{{\hat \nabla}} \def\hDelta{{\hat \Delta}} \def\hR{{\hat R}} \def\C{{\bf C}} \def\V3{{\rm V}_3} \def\bx{{\bf x}}  \def\QG{{\rm QG}}   \def\hq{\hat{q}} \def\hp{\hat{p}} \def\prm{m^\prime}

\begin{document}

\begin{titlepage}

\vspace{5mm}

\begin{center}
{\Large {\bf Vertex Operators in 4D Quantum Gravity \\ Formulated as CFT}} 
\end{center}

\vspace{5mm}

\begin{center}
{\sc Ken-ji Hamada}
\end{center}

\begin{center}
{\it Institute of Particle and Nuclear Studies, KEK, Tsukuba 305-0801, Japan} \\ and \\
{\it Department of Particle and Nuclear Physics, The Graduate University for Advanced Studies (Sokendai), Tsukuba 305-0801, Japan}
\end{center}

\begin{abstract}
We study vertex operators in 4D conformal field theory derived from quantized gravity, whose dynamics is governed by the Wess-Zumino action by Riegert and the Weyl action. Conformal symmetry is equal to diffeomorphism symmetry in the ultraviolet limit, which mixes positive-metric and negative-metric modes of the gravitational field and thus these modes cannot be treated separately in physical operators. In this paper, we construct gravitational vertex operators such as the Ricci scalar, defined as space-time volume integrals of them are invariant under conformal transformations. Short distance singularities of these operator products are computed and it is shown that their coefficients have physically correct sign. Furthermore, we show that conformal algebra holds even in the system perturbed by the cosmological constant vertex operator as in the case of the Liouville theory shown by Curtright and Thorn. 
\end{abstract}

\end{titlepage}

\section{Introduction}
\setcounter{equation}{0}
\noindent

Through many attempts quantizing gravity in four dimensions \cite{dewitt, dewitt1967, tv, veltman, weinberg, ud, stelle, tomboulis, ft}, we have come to think that a certain non-perturbative method would be necessary to formulate quantum theory of gravity in the ultraviolet (UV) limit. Since the metric field has the conformal factor, the space-time would have an exact conformal invariance when this factor fully fluctuates quantum mechanically. Conformal field theory (CFT) is thus a reliable method to describe such a quantum space-time that would be realized far beyond the Planck scale.

Recently, we proposed renormalizable 4D quantum theory of gravity formulated as a perturbed theory from such a CFT \cite{hamada02, hamada09a}, whose dynamics is described by the combined four-derivative system of the Wess-Zumino action found by Riegert \cite{riegert, am, amm92, amm97a, amm97b, hs, hh, hamada05, hamada09b} and the Weyl action. Conformal symmetry is realized as diffeomorphism symmetry in the UV limit when we quantize these actions.

In general, four-derivative quantum theories of gravity \cite{ud, stelle, tomboulis, ft} have various advantages: unlike the Einstein gravity, gravitational coupling constants become dimensionless, space-time singularities can be removed systematically, and their actions are bounded from below. But, they have ghosts. We know that in fully perturbative methods the ghost problem cannot be avoided because the ghost mode appears as a gauge invariant state in this case. Therefore, we need a non-perturbative idea to make ghosts not dynamical or not gauge invariant.

There are various proposals in the past how to treat the problem of ghosts. For instance, Tomboulis \cite{tomboulis} proposed that ghosts might be removed in the infrared (IR) region using the idea of Lee and Wick \cite{lw} based on the resummed propagator in asymptotically free theories. Recently, Horava \cite{horava} proposed a higher-derivative gravity model by making ghosts non-dynamical at the cost of diffeomorphism invariance in the UV limit. Besides, there are various approaches expanding in derivatives about the ghost-free Einstein theory with a UV cutoff. The attempt making counterterms vanish by equations of motion is this case. The idea of asymptotic safety by Weinberg \cite{weinberg} is that assuming the existence of a non-trivial UV fixed point, the cutoff is taken to be infinity at there.

Our proposal is that the problem of ghosts should be reconsidered under the light of CFT, because conformal symmetry is equal to diffeomorphism symmetry in the UV limit and mixes positive-metric and negative-metric modes of the gravitational field as
\begin{eqnarray}
   \left[ Q_M, b^\dag_{J N_1} \right]
   &=&  \sq{2J(2J+2)} \sum_{N_2} 
        \eps_{N_2} \C^{\half M}_{J N_1, J-\half -N_2} b^\dag_{J-\half N_2}
             \nonumber \\
    && - \sum_{N_2} \eps_{N_2} \C^{\half M}_{J N_1, J+\half -N_2} a^\dag_{J+\half N_2},
\end{eqnarray}
where $a^\dag_{JN}$ is the positive-metric creation mode, $b_{JN}^\dag$ is the ghost creation mode with negative-metric, and $Q_M$ is the generator of special conformal transformation given in the next section. Therefore, the ghost mode itself becomes not gauge invariant and thus we cannot consider this mode independently. This is the key of this CFT approach, not found in usual perturbative methods without such a symmetry. The field acts as a whole in physical quantities so that the correctness of the overall sign of the gravitational action becomes significant for unitarity.

In this paper, we study the physical vertex operator $\cal{O}$ that is defined as the space-time volume integral of it  satisfies the commutator
\begin{equation}
      \left[ Q_\zeta, \int d \Om_4 \cal{O} \right] =0
\end{equation}
for all generators of conformal algebra, $Q_\zeta$. Such a operator was somewhat discussed by analogy with two dimensional case in \cite{amm97b}, but how to construct it and whether it is indeed true are not clear at all. We settle the problem by applying the technique for quantum diffeomorphism developed in the previous work \cite{hamada09b} to composite fields and concretely construct the vertex operators corresponding to the cosmological term and the Ricci scalar.

Furthermore, we compute the operator products of these vertex operators and show that the coefficients of the most singular parts have a physically correct sign. In this way, we reconsider the issue of unitarity in terms of conformal field theory, namely whether correlation functions are physical or not. We see that there is no unphysical behavior of correlations at least within discussions given in this paper.

Finally, we consider the conformally invariant deformation by the cosmological constant vertex operator and its physical meanings.

\section{Diffeomorphism and Conformal Algebra}
\setcounter{equation}{0}
\noindent

In this section, we briefly review the recent results in 4D quantum gravity formulated as a certain CFT and settle the conventions and notations.

\paragraph{The Model}
The metric field is decomposed into the conformal factor and the traceless tensor field as
\begin{equation}
     g_{\mu\nu}=e^{2\phi}(\hg e^{th})_{\mu\nu}= e^{2\phi} \left( \hg_{\mu\nu} + t h_{\mu\nu} + \cdots \right)
       \label{metric decomposition}
\end{equation}
with $tr(h)=\hg^{\mu\nu}h_{\mu\nu}=0$. Here, $\hg_{\mu\nu}$ is the background metric. The conformal factor is now treated exactly without introducing its own coupling constant, while the traceless tensor field is handled perturbatively by the dimensionless coupling constant $t$.

This perturbation theory is defined by the Weyl action with the coupling constant $t$ as $I =-(1/t^2) \int d^4 x \sq{-g} C_{\mu\nu\lam\s}^2 $.\footnote{ 
The Euler term is also added in the action, while the $R^2$ term is excluded by imposing the Wess-Zumino consistency condition \cite{riegert, bcr, wz}. 
} 
Quantization is carried out by the path integral as follows: 
\begin{equation}
    \int [dg]_g \exp(iI)=\int [d\phi dh ]_\hg \exp ( iS+iI ).
\end{equation} 
Here, $S$ is the Wess-Zumino action \cite{wz} induced from diffeomorphism invariant measure $[dg]_g$ in order to preserve diffeomorphism invariance when we rewrite the path integral using the practical measure defined on the background metric $[d\phi dh]_\hg$.

In this paper we consider the UV limit of $t \to 0$ indicated by the asymptotically free behavior of the coupling. In the UV limit, the induced action $S$ is given by the Wess-Zumino action found by Riegert \cite{riegert} 
\begin{equation}
     S_{\rm RWS} = -\fr{b_1}{(4\pi)^2} \int d^4 x \sq{-\hg} \left\{ 2\phi \hDelta_4 \phi 
                   + \left( \hat{G}_4 -\fr{2}{3} \hnb^2 \hR \right) \phi \right\}, 
           \label{Riegert action}
\end{equation}
where $G_4$ is the Euler density and $\sq{-g}\Delta_4$ is the conformally invariant fourth-order differential operator. The coefficient $b_1$ has the correct sign of positive-definite such that the action is bounded from below.\footnote{ 
In general, it is given by $b_1 = 769/180 + ( N_X + 11N_W/2 + 62 N_A )/360$. Here, the first term is the loop correction from quantized gravity \cite{amm92,hs}. The other terms are the contributions from matter fields conformally coupled to gravity, where $N_X$, $N_W$, and $N_A$ are numbers of scalar fields, Weyl fermions, and gauge fields, respectively \cite{cd,ddi,duff}.
}  

In this way, the quantized Riegert-Wess-Zumino model \cite{am, amm92, amm97b, hs} originally studied based on the analogy of 2D quantum gravity is now incorporated into the renormalizable quantum gravity with only one coupling constant $t$ that indicates the asymptotic freedom \cite{hamada02, hh, hamada09a, hamada09b}. As shown in \cite{hamada09b}, quantum diffemorphism algebra in four dimensions becomes complete in the combined system of the Weyl action and the Riegert-Wess-Zumino action.\footnote{ 
Here, the $R^2$ action is not necessary to derive the generator of quantum diffeomorphism. This is also the reason why there is no $R^2$ term in the action.
} 

The Riegert-Wess-Zumino action is just 4D counter quantity of the Liouville action in 2D quantum gravity \cite{polyakov, ct, kpz, dk, seiberg}. In imitation of the Liouville field, we call the $\phi$ field in the conformal factor the Riegert field in the following.

\paragraph{Conformal Symmetry} Diffeomorphism is defined by the transformation $\dl_\xi g_{\mu\nu}=g_{\mu\lam}\nb_\nu \xi^\lam + g_{\nu\lam}\nb_\mu \xi^\lam$. Conformal symmetry is realized in the UV limit as the residual gauge symmetry of the diffeomorphism in the radiation$^+$ gauge \cite{hh, hamada09b}.\footnote{ 
The radiation gauge is the Coulomb gauge $\hnb^i h_{ij}= \hnb^i h_{i0}=0$ with $h_{00}=0$ and the superscript $+$ denotes the extra condition that the lowest mode of $h_{i0}$, satisfying $(\hnb^j \hnb_j +2)h_{i0}=0$, is removed.
} 
The gauge parameter $\xi^\mu =\zeta^\mu$ is the conformal Killing vector satisfying $\hnb_\mu \zeta_\nu + \hnb_\nu \zeta_\mu - \hg_{\mu\nu} \hnb_\lam \zeta^\lam/2 =0$ and the gauge transformation is given under the decomposition (\ref{metric decomposition}) by
\begin{equation}
    \dl_\zeta \phi = \zeta^\lam \hnb_\lam \phi + \fr{1}{4} \hnb_\lam \zeta^\lam 
        \label{conformal transformation of Riegert field}
\end{equation}
and
\begin{equation}
    \dl_\zeta h_{\mu\nu} = \zeta^\lam \hnb_\lam h_{\mu\nu} 
              + \half h_{\mu\lam} \left( \hnb_\nu \zeta^\lam - \hnb^\lam \zeta_\nu \right)
              + \half h_{\nu\lam} \left( \hnb_\mu \zeta^\lam - \hnb^\lam \zeta_\mu \right) 
         \label{conformal transformation of tensor field}
\end{equation}
at the vanishing limit of the coupling constant. This is the conformal transformation when we consider quantum gravity as a quantum field theory defined on the background $\hg_{\mu\nu}$.

The generator of $\dl_\zeta \phi$ is derived from the Riegert-Wess-Zumino action. This is one of the evidence that this action is indeed necessary to preserve diffeomorphism invariance quantum mechanically. Similarly, the generator of $\dl_\zeta h_{\mu\nu}$ is derived from the Weyl action. The right-hand side of $\dl_\zeta$ is field-dependent so that this gauge symmetry gives stringent constraints to physical quantities.

To advance the study further, we specify the background to be the cylindrical background $R\times S^3$ (\ref{metric}) as used in the previous studies. We then obtain the 15 conformal Killing vectors $\zeta^\mu= \eta^\mu, \zeta_{MN}^\mu, \zeta_M^\mu, \zeta_M^{\mu*} $ defined in Appendix. For these vectors, we obtain the 15 generators of conformal symmetry $Q_\zeta$, which are represented as follows: the Hamiltonian $H$, the 6 generators of the rotation group on $S^3$ denoted by $R_{MN}$ and the $4+4$ generators of the special conformal transformations denoted by $Q_M$ and its hermitian conjugate $Q_M^\dag$. These generators form the closed algebra of $SO(4,2)$:
\begin{eqnarray}
     \left[ Q_M, Q^\dag_N \right] &=& 2\dl_{MN} H + 2R_{MN},
           \nonumber \\
    \left[ H, Q_M \right] &=& -Q_M,
           \nonumber \\
    \left[ H, R_{MN} \right] &=& \left[ Q_M, Q_N \right] = 0,
           \nonumber  \\
    \left[ Q_M, R_{M_1 M_2} \right] &=& \dl_{M M_2}Q_{M_1}
                 -\eps_{M_1}\eps_{M_2}\dl_{M -M_1}Q_{-M_2} ,
            \nonumber  \\
    \left[ R_{M_1 M_2}, R_{M_3 M_4} \right]
        &=& \dl_{M_1 M_4} R_{M_3 M_2} 
            -\eps_{M_1}\eps_{M_2} \dl_{-M_2 M_4} R_{M_3 -M_1}
            \nonumber \\
    && - \dl_{M_2 M_3} R_{M_1 M_4} 
       +\eps_{M_1}\eps_{M_2} \dl_{-M_1 M_3} R_{-M_2 M_4} .
            \label{conformal algebra}
\end{eqnarray}
Here, the rotation generator that satisfies the relations $R_{MN}=-\eps_M \eps_N R_{-N-M}$ and $R^\dag_{MN}=R_{NM}$ forms the algebra of $SO(4)=SU(2) \times SU(2)$.

\paragraph{Explicit Form of The Generator}
We here present the notation of the field operator and the explicit form of the generator of conformal algebra \cite{amm97b, hh, hamada09b} used in this paper.

The Riegert field is decomposed into three parts: creation modes, zero-modes, and annihilation modes as
\begin{eqnarray}
    \phi = \phi_> + \phi_0  + \phi_< ,
\end{eqnarray}
where $\phi_> = \phi_<^\dag$ and
\begin{eqnarray}
   \phi_0 &=& \fr{1}{\sq{2b_1}} \left( \hq + \eta \hp \right), 
                     \nonumber \\
   \phi_< &=& \fr{\pi}{2\sq{b_1}} \biggl\{ 
               \sum_{J \geq 1/2} \sum_N \fr{1}{\sq{J(2J+1)}}a_{JN} e^{-i2J\eta}Y_{JN}
                      \nonumber \\
          && \quad
              +\sum_{J \geq 0} \sum_N \fr{1}{\sq{(J+1)(2J+1)}}b_{JN} e^{-i(2J+2)\eta}Y_{JN} \biggr\}.
\end{eqnarray}
Quantization has been carried out in the standard method such as the Dirac procedure for higher-derivative fields. The commutation relation for each mode is given by
\begin{eqnarray}
     [\hat{q}, \hat{p}] &=& i, 
             \nonumber \\
     \left[ a_{J_1 M_1},a^\dag_{J_2 M_2} \right] &=& \dl_{J_1 J_2}\dl_{M_1 M_2},
              \nonumber \\
     \left[ b_{J_1 M_1},b^\dag_{J_2 M_2} \right] &=& -\dl_{J_1 J_2}\dl_{M_1 M_2},
\end{eqnarray}
where $a_{JM}$ and $b_{JM}$ are the positive-metric and the negative-metric modes, respectively.

The generators of conformal transformations are given as follows. The Hamiltonian is
\begin{equation}
    H = \half {\hat p}^2 + b_1 + \sum_{J \geq 0} \sum_M 
           \{ 2J a^\dag_{JM} a_{JM} -(2J+2)b^\dag_{JM} b_{JM} \} ,
        \label{Hamiltonian}
\end{equation}
where the constant energy shift $b_1$ is the Casimir effect depending on the coordinate system. The generator of special conformal transformation has the form
\begin{eqnarray}
    Q_M &=& \left( \hbox{$\sq{2b_1}$}-i\hat{p} \right) a_{\half M}
             \nonumber \\
        &&  +\sum_{J \geq 0}\sum_{M_1}\sum_{M_2} \C^{\half M}_{JM_1, J+\half M_2}
              \Bigl\{ \sq{2J(2J+2)} \eps_{M_1} a^\dag_{J-M_1} a_{J+\half M_2}
                     \\
        && \quad
             -\sq{(2J+1)(2J+3)} \eps_{M_1} b^\dag_{J-M_1} b_{J+\half M_2}
             + \eps_{M_2} a^\dag_{J+\half -M_2} b_{J M_1} \Bigr\} ,
                 \nonumber 
                 \label{generator of special conformal transformation}
\end{eqnarray}
where the $\C$-function is defined by (\ref{C function}). Unlike the Hamiltonian, this generator links between the modes with different representation indices. Furthermore, as emphasized in Introduction, this generator mixes the positive-metric and the negative-metric modes due to the last cross term.

The rotation generator $R_{MN}$ is not depicted here (see refs.\cite{amm97b, hh}) because  compared with the special conformal transformation, the role of this generator is rather trivial when we construct vertex operators.

The generator of conformal transformation $\dl_\zeta h_{\mu\nu}$ (\ref{conformal transformation of tensor field}) has been constructed from the Weyl action in \cite{hh} and the transformation law was investigated in \cite{hamada09b}.  This generator also has the property that mixes positive-metric and negative-metric modes in the traceless tensor field.

In this paper, we study gravitational quantities composed of the Riegert field only such as the cosmological term and the Ricci scalar.

\section{Transformation Law of The Riegert Field}
\setcounter{equation}{0}
\noindent

We here rederive the transformation law of the Riegert field studied in ref.\cite{hamada09b}. Using this result, we study the transformation properties of various composite field operators in the half of this section and following sections.

We first see the special conformal transformation. For the creation mode $\phi_>$, we obtain
\begin{eqnarray}
     i \left[ Q_M , \phi_> \right] 
      &=& \fr{\pi}{2\sq{b_1}} i e^{i\eta} \biggl\{ 
               \left( \hbox{$\sq{2b_1}$} - i\hp \right) Y_{\half M}^*
                   \nonumber \\
     && \quad 
        + \sum_{J \geq 1/2} \sq{\fr{4J}{2J+1}}\sum_{N,N^\pp} \eps_N 
                 \C^{\half M}_{J-N,J+\half N^\pp}a^\dag_{JN} e^{i2J\eta} Y_{J+\half N^\pp}^*  
                   \nonumber \\
     && \quad 
        + \sum_{J \geq 0} \sq{\fr{2J+1}{J+1}}\sum_{N,N^\pp} \eps_N 
                 \C^{\half M}_{J-N, J+\half N^\pp} b^\dag_{JN} e^{i(2J+2)\eta} Y_{J+\half N^\pp}^*  
                   \nonumber \\
     && \quad
        - \sum_{J \geq 1/2} \fr{1}{\sq{J(2J+1)}}\sum_{N,N^\pp} \eps_N 
                 \C^{\half M}_{J-N, J-\half N^\pp}a^\dag_{JN} e^{i2J\eta} Y_{J-\half N^\pp}^*  
                 \biggr\}
                   \nonumber \\
     &=& \zeta_M^\mu \hnb_\mu \phi_> + \zeta_M^0 \pd_\eta \phi_0 + \fr{1}{4} \hnb_\mu \zeta_M^\mu . 
               \label{commutator with Q_M 1}
\end{eqnarray}
Here, in order to show the second equality, we use the product expansion
\begin{eqnarray}
    && Y_{\half M}^* Y_{JN} = \fr{1}{\sq{\V3}}
             \left\{ \sum_{N^\pp} \C^{\half M}_{JN,J+\half N^\pp}Y^*_{J+\half N^\pp}
                     + \sum_{N^\pp} \C^{\half M}_{JN,J-\half N^\pp}Y^*_{J-\half N^\pp} \right\},
                    \nonumber \\
   && \hnb^i Y_{\half M}^* \hnb_i Y_{JN} = \fr{1}{\sq{\V3}}
             \biggl\{ -2J\sum_{N^\pp} \C^{\half M}_{JN,J+\half N^\pp}Y^*_{J+\half N^\pp}
             \nonumber \\
   && \qquad\qquad\qquad\qquad\qquad
             +(2J+2) \sum_{N^\pp} \C^{\half M}_{JN,J-\half N^\pp}Y^*_{J-\half N^\pp} \biggr\} . 
      \label{scalar harmonics product expansion}
\end{eqnarray}
Similarly, we obtain the following equation
\begin{eqnarray}
   i \left[ Q_M, \phi_0 + \phi_< \right] = \zeta_M^\mu \hnb_\mu \phi_< .
          \label{commutator with Q_M 2}
\end{eqnarray}

Combining these equations, we find that the special conformal transformation of the Riegert field is written in terms of the commutator as
\begin{equation}
     i[Q_M, \phi] = \zeta_M^\mu \hnb_\mu \phi +\fr{1}{4} \hnb_\mu \zeta^\mu_M  .
\end{equation} 
The right-hand side is equal to $\dl_\zeta \phi$ (\ref{conformal transformation of Riegert field}) with the conformal Killing vector $\zeta^\mu_M$ (\ref{zeta_S vector}).

The commutators with the Hamiltonian and the rotation generator are given by 
\begin{eqnarray}
     i \left[ H, \phi \right] &=& \pd_\eta \phi ,
            \nonumber \\
     i \left[ R_{MN}, \phi \right] &=& \hnb_j \left( \zeta_{MN}^j \phi \right) ,
        \label{comutators with H and R}
\end{eqnarray}
where $\hnb_i \zeta_{MN}^i =0$ is used. Unlike the generator of special conformal transformation, these generators do not mix the zero mode and the oscillation modes.

Next, we consider the transformation law of the composite operator defined by the normal ordering,
\begin{equation}
   : \phi^n : = : \left( \phi_> + \phi_0 + \phi_< \right)^n :
              = \sum_{k=0}^n \fr{n!}{(n-k)!k!}\phi_>^{n-k} \left(\phi_0 + \phi_< \right)^k .
\end{equation}
Using (\ref{commutator with Q_M 1}) and (\ref{commutator with Q_M 2}), we find the following transformation law:
\begin{equation}
   i \left[ Q_M, :\phi^n: \right] = \zeta_M^\mu \hnb_\mu :\phi^n: + \fr{n}{4} \hnb_\mu \zeta_M^\mu :\phi^{n-1}:
     -\fr{1}{16b_1}n(n-1) \hnb_\mu \zeta^\mu_M :\phi^{n-2}: .
        \label{conformal transformation of phi^n}
\end{equation}
Here, we take account of the commutation relation $[\phi_0, \pd_\eta \phi_0 ] =i/2b_1$ such that
\begin{eqnarray}
       \pd_\eta \left( \phi_0 + \phi_< \right)^k 
       &=&  k \pd_\eta \phi_< \left( \phi_0 + \phi_< \right)^{k-1}
            +k \pd_\eta \phi_0 \left( \phi_0 + \phi_< \right)^{k-1}
               \nonumber \\
       &&  +i\fr{1}{4b_1} k(k-1) \left( \phi_0 + \phi_< \right)^{k-2}
\end{eqnarray}
and the relation $i\zeta^0_M =\hnb_\mu \zeta^\mu_M /4$.

The commutators with the Hamiltonian and the rotation generator are computed as 
\begin{eqnarray}
     i \left[ H, :\phi^n: \right] &=& \pd_\eta :\phi^n : ,
            \nonumber \\
     i \left[ R_{MN}, :\phi^n: \right] &=& \hnb_j \left( \zeta_{MN}^j  :\phi^n: \right).
\end{eqnarray}

\section{Cosmological Constant Vertex Operator}
\setcounter{equation}{0}
\noindent

Let us construct vertex operators in 4D quantum gravity. We first study the vertex operator given by the purely exponential function of the Riegert field. The normal ordering of such a composite operator is defined by
\begin{eqnarray}
    V_\a = : e^{\a \phi}: = \sum_{n=0}^\infty \fr{\a^n}{n!}:\phi^n: 
    = e^{\a \phi_>} e^{\a \phi_0} e^{\a \phi_<} .
          \label{definition of V}
\end{eqnarray}
The zero-mode part can be written as
\begin{eqnarray}
   e^{\a \phi_0} = e^{\fr{\a}{\sq{2b_1}}\hq} e^{\fr{\a}{\sq{2b_1}}\eta\hp} e^{-i\fr{\a^2}{4b_1}\eta}
                 = e^{\fr{\a}{\sq{2b_1}}\eta\hp} e^{\fr{\a}{\sq{2b_1}}\hq} e^{i\fr{\a^2}{4b_1}\eta} .
\end{eqnarray}
Here, $\a$ is a real constant, which will be determined by physical conditions given below. This constant denotes a quantum correction to the composite operator.

Using the transformation law of the composite operator given by (\ref{conformal transformation of phi^n}), we can show
\begin{eqnarray}
    i \left[ Q_M , V_\a \right] 
    = \zeta_M^\mu \hnb_\mu V_\a + \fr{h_\a}{4} \hnb_\mu \zeta_M^\mu V_\a ,
       \label{conformal transformation of V}
\end{eqnarray}
where $h_\a$ is the conformal weight of the vertex operator computed to be
\begin{eqnarray}
      h_\a = \fr{\a ( 4b_1-\a )}{4b_1},
        \label{conformal weight}
\end{eqnarray}
which satisfies the duality relation $h_\a = h_{4b_1-\a}$. This transformation law was first studied in ref.\cite{amm97b} by analogy with 2D quantum gravity, in which the equality was checked by comparing both sides level by level in the mode expansion. Here, we confirm it using the transformation property of the Riegert field.

If we take the conformal weight to be the dimension of space-time as $h_\a =4$, the transformation law (\ref{conformal transformation of V}) becomes
\begin{equation}
    i \left[ Q_M , V_\a \right] = \hnb_\mu \left( \zeta_M^\mu V_\a \right) ,
        \label{commutator of Q_M and V with weight 4}
\end{equation} 
and thus we obtain
\begin{equation}
    \left[ Q_M , \int d\Om_4 V_\a \right] =0 ,
     \label{commutator between Q_M and space-time integral of V}
\end{equation}
where $d\Om_4 = d^4 x \sq{-\hg}=d\eta d\Om_3$ is the space-time volume element.

Since the constant $\a$ satisfying the physical condition $h_\a=4$ is a real number due to $b_1 >4$ independently of matter field contents, the vertex operator $V_\a$ is real. Therefore, by taking the hermitian conjugate of (\ref{commutator between Q_M and space-time integral of V}), we obtain
\begin{equation}
    \left[ Q^\dag_M , \int d\Om_4 V_\a \right] =0 .
\end{equation}

The commutators with the Hamiltonian and the rotation generator are given by
\begin{eqnarray}
    i \left[ H, V_\a \right] &=& \pd_\eta V_\a, 
             \nonumber \\
    i \left[ R_{MN}, V_\a \right] &=& \hnb_j \left( \zeta_{MN}^j  V_\a \right) 
         \label{commutators of H or R and V}   
\end{eqnarray}
for any $\a$. Therefore, the space-time volume integral of $V_\a$ commutes with the Hamiltonian and the rotation generator.

Thus, we show that the space-time volume integral of $V_\a$ with $h_\a=4$ commutes with all generators of conformal algebra. This operator corresponds to the cosmological term.\footnote{ 
Precisely, one of the solutions of $h_\a=4$ satisfying $\a \to 4$ in the large $b_1$ limit is identified with the cosmological term, which is given by $\a=2b_1 (1-\sq{1-4/b_1})$. 
} 

\paragraph{Operator Products}
Now, we calculate the operator product of the vertex operator $V_\a$. We first compute the operator product of the Riegert field
\begin{equation}
    \phi(x) \phi (x^\pp) = \half \left[ \phi_0(\eta), \phi_0(\eta^\pp) \right] 
                          + \left[ \phi_<(x), \phi_>(x^\pp) \right] + :\phi(x) \phi(x^\pp): .
\end{equation}
The normal ordering of the product of two separate $\phi$ is defined through this equation.

The singular part can be computed using the formula \cite{vmk}
\begin{equation}
   \sum_N Y_{JN}(\bx) Y^*_{JN}(\bx^\pp) = \fr{2J+1}{V_3} \chi^J(\om) ,
\end{equation}
where $\chi^J$ is the character of the representation of $SU(2)$ group
\begin{equation}
     \chi^J (\om) = \fr{\sin [ (2J+1)\fr{\om}{2} ]}{\sin \fr{\om}{2}}
\end{equation}
and
\begin{equation}
    \cos \fr{\om}{2} = \cos \fr{\b-\b^\pp}{2} \cos \fr{\a-\a^\pp}{2} \cos \fr{\gm-\gm^\pp}{2}
                      - \cos \fr{\b+\b^\pp}{2} \sin \fr{\a-\a^\pp}{2} \sin \fr{\gm-\gm^\pp}{2} .
\end{equation}
Using this formula we obtain
\begin{eqnarray}
    && \left[ \phi_<(x), \phi_>(x^\pp) \right] 
             \nonumber \\
    && = \fr{1}{4b_1} \fr{\pi^2}{V_3} \left\{ \sum_{J \geq 1/2} \fr{1}{J}e^{-i2J(\eta-\eta^\pp)} \chi^J(\om)
         - \sum_{J \geq 0} \fr{1}{J+1} e^{-i(2J+2)(\eta-\eta^\pp)}\chi^J(\om) \right\}
             \nonumber \\
    && = \fr{1}{4b_1} \fr{\pi^2}{V_3} \sum_{J \geq 0} \fr{4}{2J+1} 
             e^{-i(2J+1)(\eta-\eta^\pp)}  \cos \left[ (2J+1)\fr{\om}{2} \right] ,
\end{eqnarray}
where the equation $\chi^{J+1/2}(\om)-\chi^{J-1/2}(\om) = 2 \cos [(2J+1)\om/2] $ is used. The infinite sum is evaluated under the regularization such as $\eta -\eta^\pp \to \eta-\eta^\pp -i\veps$, where $\veps$ is an infinitesimal positive constant. Thus, we obtain
\begin{equation}
    \left[ \phi_< (x), \phi_> (x^\pp) \right] 
    = -\fr{1}{4b_1} \log \left( 1-2e^{-i(\eta-\eta^\pp)}\cos \fr{\om}{2} + e^{-2i(\eta-\eta^\pp)} \right) .
\end{equation}

Combining this equation with the commutation relation $[ \phi_0(\eta), \phi_0(\eta^\pp)]=-i(\eta-\eta^\pp)/2b_1$, we obtain the following operator product:
\begin{equation}
   \phi(x) \phi(x^\pp) = -\fr{1}{4b_1} \log L^2(\eta-\eta^\pp,\om) + : \phi (x) \phi (x^\pp) : ,
\end{equation}
where the function $L$ is defined by
\begin{equation}
    L^2(\eta-\eta^\pp,\om) = 2 \left\{ \cos(\eta-\eta^\pp) -\cos \fr{\om}{2} \right\} .
\end{equation}
The logarithmic term gives the Green function of the Riegert field, which reflects that the field is dimensionless. 
In the short distance limit, it is 
\begin{equation}
     L^2 \simeq -(\eta-\eta^\pp)^2 + \fr{1}{4}\om^2 = (x-x^\pp)^2 ,
\end{equation}
where $\om^2 \simeq (\a-\a^\pp)^2 +(\b-\b^\pp)^2 +(\gm-\gm^\pp)^2 + 2(\a-\a^\pp)(\gm-\gm^\pp)$. This corresponds to the distance between two separate points on the flat background obtained by taking the large radius limit, and then the Green function of the Riegert field reduces to the known form of $(-1/4b_1) \times \log (x-x^\pp)^2$.

The operator product of the vertex operator $V_\a$ is calculated using the equations above to be
\begin{eqnarray}
    V_\a (x) V_{\a^\pp} (x^\pp) 
    &=& \exp \left( \a\a^\pp \left\{ \half [\phi_0(\eta), \phi_0(\eta^\pp)] 
                                     + \left[ \phi_< (x), \phi_> (x^\pp) \right] \right\} \right)
                            : V_\a (x) V_{\a^\pp} (x^\pp) : 
                   \nonumber \\
    &=& \left( \fr{1}{L^2(\eta-\eta^\pp,\om)} \right)^{\fr{\a\a^\pp}{4b_1}} 
                 : V_\a (x) V_{\a^\pp} (x^\pp) : ,        
             \label{V-V operator product}         
\end{eqnarray}
where the normal ordered product in the right-hand side is defined by
\begin{equation}
     : V_\a (x) V_{\a^\pp} (x^\pp) : = e^{\a\phi_0(\eta)+\a^\pp\phi_0(\eta^\pp)} e^{\a \phi_>(x)} e^{\a^\pp \phi_>(x^\pp)}
                                 e^{\a \phi_<(x)} e^{\a^\pp \phi_<(x^\pp)} .
\end{equation}
The zero-mode part is defined through the Baker-Campbell-Hausdorff formula $e^A e^B = e^{\half [A,B]}e^{A+B}$ satisfying when $[A,B]$ is constant.

If we take $\a$ to be a solution of $h_\a=4$ and $\a^\pp$ to be its dual solution $4b_1-\a$, we obtain the equation
\begin{eqnarray}
     V_\a (x) V_{4b_1-\a}(x^\pp) 
            &=& \left( \fr{1}{L^2(\eta-\eta^\pp,\om)} \right)^4 :V_\a (x) V_{4b_1-\a}(x^\pp):
                             \nonumber \\
            &\simeq& \fr{1}{(x-x^\pp)^8} V_{4b_1}(x) .
               \label{V-dualV operator product}
\end{eqnarray}
In the second line, we extract the most singular term in the short distance limit. The operator $V_{4b_1}$ is the dual operator of the identity with the vanishing conformal weight, $h_{4b_1}=0$.

\section{Ricci Scalar Vertex Operator}
\setcounter{equation}{0}
\noindent

Next, we consider vertex operators with derivatives. Because of the rotation invariance, the number of derivatives must be even. The simplest one is the vertex operator corresponding to the Ricci scalar, but its exact form is unknown yet. In this section, we construct the Ricci scalar vertex operator concretely using the transformation properties of the Riegert field and then study its operator product.

We here look for the vertex operator by trial and error. Considering that the Ricci scalar $\sq{-g}R$ is described by $-6 \sq{-\hg}e^{2\phi}(\hnb^2 \phi+\hnb_\mu \phi \hnb^\mu \phi -1 )$ on $R \times S^3$ classically, we first study the following normal ordered operator:
\begin{eqnarray}
   W_\b &=& : \left( \hnb^2 \phi -1 + \hnb_\mu \phi \hnb^\mu \phi \right) e^{\b \phi} :
               \nonumber \\
        &=& W_\b^{\rm A} + W_\b^{\rm B} + W_\b^{\rm C} + W_\b^{\rm D},
\end{eqnarray}
where
\begin{eqnarray}
     W_\b^{\rm A} &=& \hnb^2 \phi_> V_\b + V_\b \hnb^2 \phi_< - V_\b ,
                \nonumber \\
     W_\b^{\rm B} &=& -\pd_\eta \phi_0 \pd_\eta \phi_0 V_\b ,
                \nonumber \\
     W_\b^{\rm C} &=& -2 \pd_\eta \phi_0 \left( \pd_\eta \phi_> V_\b + V_\b \pd_\eta \phi_< \right), 
                 \nonumber \\
     W_\b^{\rm D} &=& -\pd_\eta \phi_> \pd_\eta \phi_> V_\b -2 \pd_\eta \phi_> V_\b \pd_\eta \phi_< 
                - V_\b \pd_\eta \phi_< \pd_\eta \phi_< 
                  \nonumber \\
             && + \hnb_j \phi_> \hnb^j \phi_> V_\b +2 \hnb_j \phi_> V_\b \hnb^j \phi_< 
                + V_\b \hnb_j \phi_< \hnb^j \phi_< .
\end{eqnarray}
Here, $V_\b$ is defined by (\ref{definition of V}) and the quantum correction $\b$ is a real constant determined by the physical condition. It is significant that the zero-mode $\pd_\eta \phi_0$ is introduced in the asymmetric form.

The commutators with the Hamiltonian and the rotation generator are given by
\begin{eqnarray}
    i \left[ H, W_\b \right] &=& \pd_\eta W_\b, 
             \nonumber \\
    i \left[ R_{MN}, W_\b \right] &=& \hnb_j \left( \zeta_{MN}^j  W_\b \right)     
\end{eqnarray}
for any $\b$. Here, we use (\ref{commutators of H or R and V}) and equations derived from (\ref{comutators with H and R}) such as $i[R_{MN}, \hnb^2 \phi ]= \hnb_j (\zeta^j_{MN}\hnb^2 \phi)$.

Let us show that this operator has a good behavior under the $Q_M$ transformation. The transformation law of each part is computed as 
\begin{eqnarray}
   i \left[ Q_M , W_\b^{\rm A} \right] &=& 
        \zeta_M^\mu \hnb_\mu W_\b^{\rm A} + \fr{h_\b +2}{4} \hnb_\mu \zeta_M^\mu W_\b^{\rm A}
        - 2 \zeta_M^0 \pd_\eta \phi_0 V_\b 
            \nonumber \\  
       && - 2 \zeta_M^0 \left( \pd_\eta \phi_> V_\b + V_\b \pd_\eta \phi_< \right)
          + 2 \zeta_M^j \left( \hnb_j \phi_> V_\b + V_\b \hnb_j \phi_< \right) ,
                \nonumber \\
   i \left[ Q_M , W_\b^{\rm B} \right] &=&       
         \zeta_M^\mu \hnb_\mu W_\b^{\rm B} + \fr{h_\b}{4} \hnb_\mu \zeta_M^\mu W_\b^{\rm B} ,
               \nonumber \\
   i \left[ Q_M , W_\b^{\rm C} \right] &=& 
          \zeta_M^\mu \hnb_\mu W_\b^{\rm C} + \fr{h_\b +1}{4} \hnb_\mu \zeta_M^\mu W_\b^{\rm C} 
           + \half \hnb_\mu \zeta_M^\mu W_\b^{\rm B}    
                    \nonumber \\
        && + 2 \zeta_M^0 \pd_\eta \phi_0 V_\b 
           - 2i \pd_\eta \phi_0 \zeta_M^j  \left( \hnb_j \phi_> V_\b + V_\b \hnb_j \phi_< \right) ,
                \nonumber \\   
   i \left[ Q_M , W_\b^{\rm D} \right] &=& 
         \zeta_M^\mu \hnb_\mu W_\b^{\rm D} + \fr{h_\b +2}{4} \hnb_\mu \zeta_M^\mu W_\b^{\rm D} 
         + \fr{1}{4} \hnb_\mu \zeta_M^\mu W_\b^{\rm C} 
              \nonumber \\
        &&  + 2 \zeta_M^0 \left( \pd_\eta \phi_> V_\b + V_\b \pd_\eta \phi_< \right)
            + 2i \pd_\eta \phi_0 \zeta_M^j \left( \hnb_j \phi_> V_\b + V_\b \hnb_j \phi_< \right)
               \nonumber \\
        &&  - 2 \zeta_M^j \left( \hnb_j \phi_> V_\b + V_\b \hnb_j \phi_< \right) ,
\end{eqnarray}
where we use equations derived from (\ref{commutator with Q_M 1}) and (\ref{commutator with Q_M 2}) and the transformation law of $V_\b$ (\ref{conformal transformation of V}). Combining these equations, we find
\begin{eqnarray}
    i \left[ Q_M , W_\b \right] 
    = \zeta_M^\mu \hnb_\mu W_\b + \fr{h_\b+2}{4} \hnb_\mu \zeta_M^\mu W_\b 
      \label{commutator of Q_M and W}
\end{eqnarray}
for any $\b$, where $h_\b$ is defined by (\ref{conformal weight}). Here, note that this equation is so strong as to determine the form of the vertex operator uniquely to be above.

Unlike $V_\b$, the vertex operator $W_\b$ is not real because it has the asymmetric form regarding to the zero-mode so that the commutator between $Q_M^\dag$ and $W_\b$ is not equal to the right-hand side of (\ref{commutator of Q_M and W}) with $\zeta^{\mu *}_M$ instead of $\zeta^\mu_M$. However, $W_\b$ and $W_\b^\dag$ are equivalent up to time derivative of $V_\b$ as
\begin{equation}
    W_\b = W_\b^\dag + \fr{i}{b_1} \pd_\eta V_\b ,
\end{equation}
where 
\begin{equation}
      \pd_\eta V_\b = \b \left\{ 
           \half \left( \pd_\eta \phi_0 V_\b + V_\b \pd_\eta \phi_0 \right) 
                  + \pd_\eta \phi_> V_\b + V_\b \pd_\eta \phi_<  \right\} .
\end{equation}
Therefore, we find that the real vertex operator defined by
\begin{equation}
        {\cal R}_\b = \half \left( W_\b + W_\b^\dag \right) 
\end{equation} 
satisfies the relation
\begin{equation}
     \int d\Om_4 {\cal R}_\b = \int d \Om_4 W_\b = \int d \Om_4 W_\b^\dag .
\end{equation}

Thus, we can show that the space-time volume integral of $\cal{R}_\b$ with $h_\b =2$ commutes with all generators of conformal algebra as
\begin{equation}
     \left[ Q_\zeta, \int d\Om_4 {\cal R}_\b \right] =0 .
\end{equation}
This vertex operator corresponds to the Ricci scalar.

We here mention that we cannot define the naive state-operator correspondence as in 2D CFT for the Ricci scalar vertex operator, although the cosmological constant vertex operator has such a correspondence.\footnote{ 
The physical states is defined by the conditions $(H-4)|{\rm phys}\rangle = R_{MN}|{\rm phys}\rangle =Q_M |{\rm phys}\rangle=0$ \cite{hamada05, hamada09b}. The state for the cosmological constant denoted by $|V_\a \rangle=e^{\a\phi_0(0)}|\Om\rangle$ can be obtained from the vertex operator $V_\a$ through the state-operator correspondence: $|V_\a \rangle = \lim_{\eta \to i\infty}e^{-4i\eta} V_\a(\eta,\bx)|\Om\rangle$, where $h_\a=4$ and $|\Om \rangle$ is the conformally invariant vacuum.
} 
It seems to be a general feature that any vertex operators with derivatives have. This issue remains in the future.

\paragraph{Another Expression}
We here consider more simple expression of the vertex operator with two derivatives defined by 
\begin{equation}
     {\cal S}_\b =  \hnb^2 \phi_> V_\b + V_\b \hnb^2 \phi_< - \fr{4}{\b} V_\b .
\end{equation}
This vertex operator is the same to $W^{\rm A}_\b$ up to the last term. We find that this operator has the following transformation property:
\begin{eqnarray}
     i \left[ Q_M, {\cal S}_\b \right] 
       &=& \zeta^\mu_M \hnb_\mu {\cal S}_\b 
       + \fr{h_\b +2}{4} \hnb_\mu \zeta^\mu_M {\cal S}_\b
            \nonumber \\
       && +\fr{2}{\b} \left\{ -\pd_\eta \left( \zeta^0_M V_\b \right) + \hnb_j \left( \zeta^j_M V_\b \right)
               + \fr{2-h_\b}{4} \hnb_\mu \zeta^\mu_M V_\b  \right\} . 
             \nonumber \\ 
       && \label{commutator of Q_M and another R}
\end{eqnarray}
Thus, we obtain the equation
\begin{equation}
       \left[ Q_M, \int d\Om_4 {\cal S}_\b \right] =0 
\end{equation} 
for $h_\b =2$.

Since this vertex operator is real, the same equation holds when we replace $Q_M$ with its hermitian conjugate $Q_M^\dag$. The commutators with the Hamiltonian and the rotation generator also vanish. Therefore, the space-time volume integral of ${\cal S}_\b$ with $h_\b =2$ commutes with all generators of conformal algebra. This operator is another expression of the physical vertex operator with two derivatives.

\paragraph{Operator Products}
Let us consider the operator product of the real vertex operator ${\cal S}_\b$. Here, we are interested in the short distance singularity. The most-singular (M.S.) part is computed as
\begin{eqnarray}
     && {\cal S}_\b (x) {\cal S}_{\b^\pp}(x^\pp)|_{\rm M.S.} 
           \nonumber \\
     && = \left\{ 
        \left[ \hnb^2 \phi_< (x), \hnb^2 \phi_> (x^\pp) \right]
        + \left[ \b \phi_< (x), \hnb^2 \phi_> (x^\pp) \right] \left[ \hnb^2 \phi_< (x), \b^\pp \phi_> (x^\pp) \right]
         \right\} 
             \nonumber \\
     && \quad \times
         \left( \fr{1}{L^2(\eta-\eta^\pp,\om)} \right)^{\fr{\b\b^\pp}{4b_1}}:V_\b (x) V_{\b^\pp}(x^\pp): ,
          \label{product expression of R}
\end{eqnarray}
where we use the operator product (\ref{V-V operator product}).

The commutators in the right-hand side can be computed using the mode-expansion
\begin{eqnarray}
   \hnb^2 \phi_< &=& \fr{2\pi}{\sq{b_1}} \biggl\{ 
                   -\sum_{J \geq 1/2} \sum_N \sq{\fr{J}{2J+1}}a_{JN}e^{-i2J\eta}Y_{JN}
                     \nonumber \\
                 && \qquad\quad
                   +\sum_{J \geq 0} \sum_N \sq{\fr{J+1}{2J+1}}b_{JN}e^{-i(2J+2)\eta}Y_{JN}  \biggr\} ,
\end{eqnarray}
and we obtain the following short distance singularities:
\begin{eqnarray}
    \left[ \hnb^2 \phi_< (x), \phi_> (x^\pp) \right] 
     &=& -\fr{1}{b_1} \fr{\cos \fr{\om}{2}}{L^2(\eta-\eta^\pp,\om)}
             \nonumber \\
     &\simeq&  -\fr{1}{b_1}\fr{1}{(x-x^\pp)^2} 
         \label{commutator of phi and d^2 phi}
\end{eqnarray}
and 
\begin{eqnarray}
    \left[ \hnb^2 \phi_< (x), \hnb^2 \phi_> (x^\pp) \right] 
    &=& \fr{8}{b_1} \fr{\cos(\eta-\eta^\pp)\cos \fr{\om}{2} -1}{ L^4(\eta-\eta^\pp,\om) }
              \nonumber \\
    &\simeq& \fr{8}{b_1} \fr{-(\eta-\eta^\pp)^2}{(x-x^\pp)^4} -\fr{4}{b_1}\fr{1}{(x-x^\pp)^2}
             \to 0 .
         \label{commutator of d^2 phi and d^2 phi}
\end{eqnarray}
The last arrow in (\ref{commutator of d^2 phi and d^2 phi}) denotes that in the flat background by the large radius limit, this commutator vanishes in proportion to the inverse of the square of radius by dimensional analysis.

The most singular part thus comes from the second term in the right-hand side of (\ref{product expression of R}). Using (\ref{commutator of phi and d^2 phi}), it is given by
\begin{eqnarray}
    {\cal S}_\b (x) {\cal S}_{\b^\pp}(x^\pp)|_{\rm M.S.} 
    &=& \fr{\b\b^\pp}{b_1^2} \fr{\cos^2 \fr{\om}{2}}{ \{ L^2(\eta-\eta^\pp,\om) \}^{2+\fr{\b\b^\pp}{4b_1}}}
      :V_\b (x) V_{\b^\pp}(x^\pp): 
         \nonumber \\
    &\simeq& \fr{\b\b^\pp}{b_1^2} \left( \fr{1}{(x-x^\pp)^2} \right)^{2+\fr{\b\b^\pp}{4b_1}} 
             :V_{\b+\b^\pp}(x): .
\end{eqnarray}
If we take $\b$ to be a solution of $h_\b =2$ and $\b^\pp$ to be its dual solution $4b_1-\b$, we obtain
\begin{equation}
    {\cal S}_\b (x) {\cal S}_{4b_1-\b}(x^\pp)|_{\rm M.S.} 
    \simeq \fr{8}{b_1} \fr{1}{(x-x^\pp)^8} V_{4b_1}(x) ,     
\end{equation}
where $V_{4b_1}$ is the identity operator with vanishing conformal weight as mentioned before.

The operator product has the physical coefficient of positive due to $b_1 >0$, as expected from the property that the vertex operator is real. The positivity of $b_1$ represents that the Riegert-Wess-Zumino action has the right sign such that the path integral is well-defined.\footnote{ 
In terms of the Wick-rotated Euclidean theory, it is simply denoted that the path integral has the correct weight $e^{-S_{\rm E}}$ with the action $S_{\rm E}$ bounded from below. 
} 
This suggests that the correctness of the overall sign of the gravitational action, not the sign of each mode, is significant for unitarity.

In general, diffeomorphism invariant fields are real and thus their correlation functions are expected to be physical as long as diffeomorphism invariance holds.

\section{Conformally Invariant Deformation}
\setcounter{equation}{0}
\noindent

Finally, we consider a deformation of the generator by a mass scale while the conformal algebra holds, as discussed by Curtright and Thorn in the Liouville theory \cite{ct, seiberg}.

We here consider the Riegert-Wess-Zumino action perturbed by the cosmological constant vertex operator as
\begin{equation}
    \bar{S}_{\rm RWZ} = S_{\rm RWZ} - \Lam \int d\Om_4 V_\a ,
     \label{deformed Riegert action}
\end{equation}
where $h_a=4$. Since the stress-tensor for the cosmological term is proportional to the background metric as $\hat{T}^\Lam_{\mu\nu} = -\Lam \hg_{\mu\nu} V_\a$, the generator is expressed as
\begin{equation}
    \bar{Q}_\zeta = Q_\zeta + \Lam  \int d\Om_3 \zeta^0 V_\a .
\end{equation}
Substituting the expressions of the conformal Killing vectors $\zeta^\mu$, we obtain
\begin{eqnarray}
    \bar{H} &=& H + \Lam  \int d\Om_3 V_\a, 
          \nonumber \\
    \bar{Q}_M &=& Q_M + \Lam  \int d\Om_3 \zeta^0_M V_\a ,
          \nonumber \\
    \bar{R}_{MN} &=& R_{MN} .
\end{eqnarray}
In the following, we will show that the generator $\bar{Q}_\zeta$ also satisfies the conformal algebra of $SO(4,2)$.

First, we see
\begin{eqnarray}
   \left[ \bar{H}, \bar{Q}_M \right]
   &=& \left[ H, Q_M \right] + \Lam \left[ H, \int d\Om_3 \zeta^0_M V_\a \right]
          \nonumber \\
   &&  - \Lam \left[ Q_M, \int d\Om_3 V_\a \right] 
       + \Lam^2 \left[ \int d\Om_3 V_\a, \int d\Om_3 \zeta^0_M V_\a \right]
          \nonumber \\
   &=& -Q_M + \Lam \left[ H, \int d\Om_3 \zeta^0_M V_\a \right]
            - \Lam \left[ Q_M, \int d\Om_3 V_\a \right] ,
\end{eqnarray}
where we use $[H,Q_M]=-Q_M$ and $[ V_\a(x), V_\a(x^\pp) ]=0$ as a consequence of the operator product (\ref{V-V operator product}). Using (\ref{commutator of Q_M and V with weight 4}) and (\ref{commutators of H or R and V}), we obtain
\begin{equation}
       \left[ \bar{H}, \bar{Q}_M \right] 
       = -Q_M -\Lam \int d\Om_3 \zeta^0_M V_\a = -\bar{Q}_M .
\end{equation}
The commutator of $\bar{H}$ and $\bar{R}_{MN}$ vanishes as
\begin{equation}
      \left[ \bar{H}, \bar{R}_{MN} \right] 
       = i \Lam \int d\Om_3 \hnb_j \left( \zeta^j_{MN}V_\a \right) = 0 ,
\end{equation}
where (\ref{commutators of H or R and V}) is used.

The commutator of $\bar{Q}_M$ and $\bar{Q}_N^\dag$ are computed as follows:
\begin{eqnarray}
      \left[ \bar{Q}_M, \bar{Q}^\dag_N \right]
       &=& \left[ Q_M, Q^\dag_N \right] + \Lam \left[ Q_M, \int d\Om_3 \zeta^{0*}_N V_\a \right] 
          - \Lam \left[ Q_N^\dag, \int d\Om_3 \zeta^0_M V_\a \right] 
            \nonumber \\
       &=& 2 \dl_{MN} H +2 R_{MN} + 2 \dl_{MN}\Lam \int d\Om_3 V_\a 
            \nonumber \\
       &=& 2\dl_{MN} \bar{H} + 2 \bar{R}_{MN} .
\end{eqnarray}
Here, we use the equation
\begin{equation}
    -\zeta^j_M \hnb_j \zeta^{0*}_N  + \pd_\eta \zeta^0_M \zeta^{0*}_N 
    = i \fr{\V3}{4} \left( \hnb^j Y^*_{\half M} \hnb_j Y_{\half N} + Y^*_{\half M} Y_{\half N}  \right)
    = i \dl_{MN} 
        \label{scalar product for commutator of Q and Q^dag}
\end{equation}
and the complex conjugate of this equation, which can be derived from (\ref{scalar harmonics product expansion}). We also find that the commutator of two $\bar{Q}_M$ vanishes as
\begin{equation}
      \left[ \bar{Q}_M, \bar{Q}_N \right]
       = \Lam \left[ Q_M, \int d\Om_3 \zeta^0_N V_\a \right] 
          - \Lam \left[ Q_N, \int d\Om_3 \zeta^0_M V_\a \right] 
       = 0 ,
\end{equation}
using (\ref{commutator of Q_M and V with weight 4}) and the relations $\pd_\eta \zeta^0_M = i \zeta^0_M$ and $\hnb^j \zeta^0_M = i \zeta^j_M$ from the definition (\ref{zeta_S vector}).

Lastly, we calculate the commutator of $\bar{Q}_M$ and $\bar{R}_{MN}$, which is given by
\begin{eqnarray}
        \left[ \bar{Q}_M, \bar{R}_{M_1 M_2} \right]
        &=& \left[ Q_M, R_{M_1 M_2} \right] 
            -\Lam \left[ R_{M_1 M_2}, \int d\Om_3 \zeta^0_M V_\a \right]
               \nonumber \\
        &=& \dl_{MM_2}Q_{M_1}-\eps_{M_1}\eps_{M_2} \dl_{M-M_1}Q_{-M_2} 
            - i\Lam \int d\Om_3 \zeta_{M_1 M_2}^j \hnb_j \zeta^0_M V_\a
               \nonumber \\
        &=& \dl_{MM_2}\bar{Q}_{M_1}-\eps_{M_1}\eps_{M_2} \dl_{M-M_1}\bar{Q}_{-M_2} ,   
\end{eqnarray} 
where the second equality comes from (\ref{commutators of H or R and V}) and the third equality is from the equation
\begin{equation}
    \zeta^j_{M_1 M_2} \hnb_j \zeta^0_M 
    = i \dl_{M M_2} \zeta^0_{M_1} - i \eps_{M_1}\eps_{M_2}\dl_{M-M_1} \zeta^0_{-M_2},
\end{equation}
which is shown by using the product formula $\hnb_i Y^*_{1/2 M}\hnb^i Y_{1/2 N} = 4\dl_{MN}/\V3 -Y^*_{1/2 M} Y_{1/2 N}$ and $Y^*_{1/2 N}=\eps_N Y_{1/2 -N}$.

Since the rotation generator is $\bar{R}_{MN}=R_{MN}$, the closed algebra of this generator is trivially satisfied. Thus, we show that the deformed generator $\bar{Q}_\zeta$ also forms the closed algebra of conformal symmetry equivalent to that of $Q_\zeta$.

This result reflects that the conformal symmetry is just equal to the diffeomorphism invariance in the UV limit. Thus, 4D quantum gravity with the cosmological term (\ref{deformed Riegert action}) also has the conformal invariance as a realization of background metric independence, and it is expected that the same situation will be realized in the system perturbed by the Ricci scalar.

\section{Conclusion and Discussion}
\setcounter{equation}{0}
\noindent

We have studied various vertex operators in 4D CFT obtained by quantizing gravity in the non-perturbative manner, whose dynamics is described by the combined system of the Riegert-Wess-Zumino action and the Weyl action. The conformal invariance is equal to the diffeomorphism invariance in the UV limit, which gives the first example of diffeomorphism algbera that is closed quantum mechanically in four dimensions.

Since the conformal symmetry mixes the positive-metric and the negative-metric modes in the gravitational field, we cannot treat these modes separately and thus the field acts as a whole when we consider physical quantities. In this paper, developing the study of the transformation law of the gravitational field achieved in our previous work \cite{hamada09b}, we have constructed the gravitational vertex operator composed of the field such as the Ricci scalar.

The short distance singularity of the product of the vertex operator has been computed and it was shown that the coefficient has the physical sign of positive when $b_1 >0$. This result seems to be natural because the positivity of two-point functions originates from the property that the vertex operator is real due to diffeomorphism invariance, and why the positivity does not break in the correlation, after all, comes from the fact that the action is bounded from below so that the path integral is well-defined.

Of course, this statement is based on the fact that physical vertex operators are written in terms of the field itself and thus the ghost mode never appears individually. If ghosts themselves were gauge invariant as in usual perturbative approaches, one could not apply this statement to their correlations because only the ghost part in the action contributes to the path integral so that the positivity of the overall sign of the action becomes meaningless.

Finally, we showed that the generator of conformal transformation perturbed by the cosmological constant vertex operator with mass scale also forms the conformal algebra. This indicates that correlation functions in such a perturbed system have a power-law behavior of this mass scale, as in the case of 2D quantum gravity \cite{kpz, dk, seiberg}.\footnote{ 
The IR scale indicating a logarithmic violation of conformal symmetry is given by the dynamical energy scale $\Lam_\QG$, and hence a degree of deviation from CFT is measured by the running coupling constant. If we take $\Lam_\QG \simeq 10^{17}$GeV below the Planck scale, we can construct the inflation model in which the Riegert field serves for the inflaton field \cite{hy, hhy06, hhy09}. 
} 


\appendix

\section{Scalar Harmonics and Conformal Killing Vectors on $R \times S^3$}
\setcounter{equation}{0}
\noindent

The background metric is parameterized by the coordinate $x^\mu=(\eta,x^i)$ using the Euler angles $x^i=(\a,\b,\gm)$ as 
\begin{equation}
     d{\hat s}^2_{R\times S^3}= \hg_{\mu\nu}dx^\mu dx^\nu 
                       = -d\eta^2 + \fr{1}{4} (d\a^2 +d\b^2 +d\gm^2 +2 \cos \b d\a d\gm ) ,
           \label{metric}
\end{equation}
where $\a$, $\b$ and $\gm$ have the ranges $[0,2\pi]$, $[0,\pi]$ and $[0,4\pi]$, respectively. The radius of $S^3$ is taken to be unity. The curvatures are then given by $\hR_{ij}=2\hg_{ij}$, $\hR=6$ and $\hR_{0\mu\nu\lam}=\hR_{0\mu}=\hat{C}^2_{\mu\nu\lam\s}=\hat{G}_4=0$. The volume element on the unit $S^3$ is $d\Om_3 = \sin \b d\a d\b d\gm/8$ and the volume is given by $\V3 = \int_{S^3} d\Om_3 =2\pi^2$.

The conformal Killing vector and the Riegert field are written in terms of scalar harmonics on $S^3$. It, denoted by $Y_{JM}$, is the eigenfunction of the Laplacian $\Box_3 =\hnb^i \hnb_i$ on $S^3$ belonging to the $(J,J)$ representation of the isometry group $SU(2)\times SU(2)$, 
\begin{equation}
         \Box_3 Y_{J M} = -2J(2J+2) Y_{J M},  \qquad  Y_{JM} = \sq{\fr{2J+1}{\V3}}D^J_{m\prm}.
           \label{scalar harmonics}
\end{equation}
Here, $D^J_{m\prm}$ is the Wigner $D$-function \cite{vmk}. $J~(\geq 0)$ takes integer or half-integer values, and the index $M=(m,\prm)$ denotes the multiplicity of the $(J,J)$ representation and thus $m$ and $\prm$ take values from $-J$ to $J$, respectively. The normalization is taken to be $\int d\Om_3 Y^*_{J_1 M_1}Y_{J_2 M_2} = \dl_{J_1J_2}\dl_{M_1 M_2}$, where $\dl_{M_1 M_2}=\dl_{m_1 m_2}\dl_{\prm_1 \prm_2}$. The complex conjugate of scalar harmonics is given by $Y^*_{J M}= \eps_M Y_{J -M}$, where the sign factor is $\eps_M=(-1)^{m-\prm}$ satisfying $\eps_M^2=1$.

The $SU(2)\times SU(2)$ Clebsch-Gordan coefficient defined by the integral of three products of scalar harmonics over $S^3$ is given by
\begin{eqnarray}
     \C^{JM}_{J_1M_1,J_2M_2}
      &=& \sq{\V3} \int_{S^3} d\Om_3 Y^*_{JM}Y_{J_1M_1}Y_{J_2M_2}
             \nonumber \\
      &=& \sq{\fr{(2J_1+1)(2J_2+1)}{2J+1}} C^{Jm}_{J_1m_1,J_2m_2}
                           C^{J\prm}_{J_1\prm_1,J_2\prm_2} ,
        \label{C function}
\end{eqnarray} 
where $C^{Jm}_{J_1 m_1, J_2 m_2}$ is the standard $SU(2)$ Clebsch-Gordan coefficient \cite{vmk}.

The 15 conformal Killing vectors on $R \times S^3$ are as follows. The Killing vector that generates the time translation is given by $\eta^\mu=(1,0,0,0)$. The $6$ Killing vectors on $S^3$ are given by $\zeta_{MN}^\mu=(0,\zeta_{M N}^j)$ with
\begin{equation}
     \zeta^j_{MN} 
      = i \fr{\V3}{4} \left\{ Y^*_{\half M} \hnb^j Y_{\half N} 
              - Y_{\half N} \hnb^j Y^*_{\half M}  \right\} ,
\end{equation}
satisfying the equation $\hnb^i \zeta^j_{MN} + \hnb^j \zeta^i_{MN}=0$. Here, we use the index without $J$ in the case of the 4-vector index with $J=1/2$ that appears in the conformal killing vectors and the corresponding generators. The $4+4$ vectors that generate the special conformal transformation are given by $\zeta_M^\mu=(\zeta_M^0,\zeta_M^j)$ with
\begin{equation} 
   \zeta^0_M = \fr{\sq{\V3}}{2} e^{i\eta}Y^*_{\half M} , \quad 
   \zeta^j_M = -i\fr{\sq{\V3}}{2} e^{i\eta} \hnb^j Y^*_{\half M} 
          \label{zeta_S vector}
\end{equation}
and their complex conjugates.



\begin{thebibliography}{99}

\bibitem{dewitt} 
B. DeWitt, in Relativity, Groups and Topology, eds. B. DeWitt and C. DeWitt (Gordon and Breach, New York, 1964).
\bibitem{dewitt1967}
B. DeWitt,  Phys. Rev. {\bf 160} (1967) 1113; Phys. Rev. {\bf 162} (1967) 1195, 1239.
\bibitem{tv} 
G. {}'t Hooft and M. Veltman, Ann. Inst. Henri Poincare {\bf XX} (1974) 69. 
\bibitem{veltman}
M. Veltman, {\it Methods in Field Theory}, Les Houches 1975 (North-Holland, Amsterdam, 1976). 
\bibitem{weinberg}
S. Weinberg, in {\it Understanding the Fundamental Constituents of Matter}, ed. A. Zichichi (Plenum Press, NY, 1977).
\bibitem{ud}
R. Utiyama and B. DeWitt, J. Math. Phys. {\bf 3} (1962) 608.  
\bibitem{stelle}
K. Stelle, Phys. Rev. {\bf D16} (1977) 953.
\bibitem{tomboulis}
E. Tomboulis, Phys. Lett. {\bf 70B} (1977) 361.
\bibitem{ft}
E. Fradkin and A. Tseytlin, Nucl. Phys. {\bf B201} (1982) 469.  
\bibitem{hamada02}
K. Hamada, Prog. Theor. Phys. {\bf 108} (2002) 399.   
\bibitem{hamada09a}
K. Hamada, Found. Phys. {\bf 39} (2009) 1356.
\bibitem{riegert}
R. Riegert, Phys. Lett. {\bf 134B} (1984) 56.
\bibitem{am}  
I. Antoniadis and E. Mottola, Phys. Rev. {\bf D45} (1992) 2013.
\bibitem{amm92}
I. Antoniadis, P. Mazur and E. Mottola, Nucl. Phys. {\bf B388} (1992) 627. 
\bibitem{amm97a}
I. Antoniadis, P. Mazur and E. Mottola, Phys. Rev. {\bf 55} (1997) 4756.
\bibitem{amm97b}
I. Antoniadis, P. Mazur and E. Mottola, Phys. Rev. {\bf 55} (1997) 4770.
\bibitem{hs}
K. Hamada and F. Sugino, Nucl. Phys. {\bf B553} (1999) 283. 
\bibitem{hh}
K. Hamada and S. Horata, Prog. Theor. Phys. {\bf 110} (2003) 1169.
\bibitem{hamada05}
K. Hamada, Int. J. Mod. Phys. {\bf A20} (2005) 5353.
\bibitem{hamada09b}
K. Hamada, Int. J. Mod. Phys. {\bf A24} (2009) 3073.
\bibitem{lw}
T. Lee and G. Wick, Nucl. Phys. {\bf B9} (1969) 209.
\bibitem{horava}
P. Horava, Phys. Rev. {\bf D79} (2009) 084008.
\bibitem{bcr}
L. Bonora, P. Cotta-Ramusino and C. Reina, Phys. Lett. {\bf B126} (1983) 305.
\bibitem{wz}
J. Wess and B. Zumino, Phys. Lett. {\bf 37B} (1971) 95.
\bibitem{polyakov}
A. Polyakov, Phys. Lett. {\bf 103B} (1981) 207.
\bibitem{ct}
T. Curtright and C. Thorn, Phys. Rev. Lett. {\bf 48} (1982) 1309.
\bibitem{kpz}
V. Knizhnik, A. Polyakov and A. Zamolodchikov, Mod. Phys. Lett. {\bf A3} (1988) 819.
\bibitem{dk}
J. Distler and H. Kawai, Nucl. Phys. {\bf B321} (1989) 509; F. David, Mod. Phys. Lett. {\bf A3} (1988) 1651.
\bibitem{seiberg}
N. Seiberg, Prog. Theor. Phys. Suppl. {\bf 102} (1990) 319.
\bibitem{cd}
D. Capper and M. Duff, Nuovo Cimento {\bf 23A} (1974) 173.
\bibitem{ddi}
S. Deser, M. Duff and C. Isham, Nucl. Phys. {\bf B111} (1976) 45.
\bibitem{duff}
M. Duff, Nucl. Phys. {\bf B125} (1977) 334.
\bibitem{vmk}
D. Varshalovich, A. Moskalev and V. Khersonskii, {\it Quanyum Theory of Angular Momentum} (World Scientific, Singapore, 1988).
\bibitem{hy}
K. Hamada and T. Yukawa, Mod. Phys. Lett. {\bf A20} (2005) 509. 
\bibitem{hhy06}
K. Hamada, S. Horata and T. Yukawa, Phys. Rev. {\bf D74} (2006) 123502.
\bibitem{hhy09}
K. Hamada, S. Horata and T. Yukawa, Phys. Rev. {\bf D81} (2010) 083533.

\end{thebibliography}
\end{document}